\documentclass[a4paper,fleqn,usenatbib]{mnras}

\usepackage{newtxtext,newtxmath}
\usepackage{booktabs,tabularx}
\usepackage{float}  

\usepackage[T1]{fontenc}
\usepackage{ae,aecompl}

\usepackage{graphicx}	
\usepackage{amsmath}	
\usepackage{amssymb}	
\usepackage[compatibility=false]{caption}
\usepackage{subcaption}
\usepackage{cleveref}

\title[Horseshoe Co-orbitals of Earth: New Candidates]{Horseshoe Co-orbitals of Earth: Current
Population and New Candidates}

\author[M. Kaplan]{
Murat Kaplan,$^{1}$\thanks{E-mail: muratkaplan@akdeniz.edu.tr}
Sergen Cengiz,$^{1}$
\\
$^{1}$Akdeniz University, Department of Space Sciences and Technologies,
Antalya, Turkey\\
}

\date{Accepted 2020 June 23. Received 2020 June 22; in original form 2019 May 07}

\pubyear{2020}

\begin{document}
\label{firstpage}
\pagerange{\pageref{firstpage}--\pageref{lastpage}}
\maketitle

\begin{abstract}
	Most co-orbital objects in the Solar system are thought to follow
	tadpole-type orbits, behaving as Trojans.  However, most of Earth's
	identified co-orbitals are moving along horseshoe-type orbits.  The
	current tally of minor bodies considered to be Earth co-orbitals amounts
	to 18; of them, 12 are horseshoes, five are quasi-satellites, and one is
	a Trojan. The semimajor axis values of all these bodies librate between $0.983$ au and
	$1.017$ au.
	In this work, we have studied the
	dynamical behaviour of objects following orbits with semimajor axis
	within this range that may be in a 1:1 mean-motion resonance with Earth.
	Our results show that asteroids 2016 CO$_{246}$, 2017 SL$_{16}$, and
	2017 XQ$_{60}$ are moving along asymmetrical horseshoe-type orbits; the
	asteroid 2018 PN$_{22}$ follows a nearly symmetric or regular
	horseshoe-type orbit. Asteroids 2016 CO$_{246}$, 2017 SL$_{16}$, and
	2017 XQ$_{60}$ can remain in the horseshoe co-orbital state for about
	$900$ yr, $3300$ yr and $2700$ yr, respectively. 
	Asteroid 2018 PN$_{22}$ has a more chaotic dynamical
	behaviour; it may not stay in a horseshoe co-orbital state for more than
	200 yr. The horseshoe libration periods of 2016 CO$_{246}$, 2017
	SL$_{16}$, 2017 XQ$_{60}$, and 2018 PN$_{22}$ are 280, 255, 411, and 125
	yr, respectively.

\end{abstract}

\begin{keywords}
celestial mechanics -- asteroids: general -- methods: numerical --- minor
	planets, asteroids: individual: 2016 CO$_{246}$ --- asteroids:
	individual: 2017 SL$_{16}$ --- asteroids: individual: 2017
	XQ$_{60}$ ---  asteroids: individual: 2018 PN$_{22}$
\end{keywords}



\section{Introduction}

Co-orbitals are in a 1:1 mean-motion resonance with the host body; they are
classified in 3 different classes according to the shape of the orbit: horseshoe
(HS), quasi-satellite (QS) or Trojan following a tadpole path (TP). In the Solar 
system, the HS-type orbits are less common than the
TP-type orbits \citep{christou2011long}.  The current tally of known co-orbitals
in the Solar system amounts to over $7,000$ (JPL's
SBDB\footnote{\url{https://ssd.jpl.nasa.gov/sbdb\_query.cgi}}); the majority of them
are Jupiter Trojans \citep{de2014comparative}. However, most known Earth co-orbitals (12 out of 18) are in the HS co-orbital state \citep{de2016asteroid}. Five of the Earth co-orbitals
follow QS-type orbits, and only one asteroid (2010 TK$_7$) follows a TP-type orbit.

In addition to these, there are transitions between the existing HS and QS	
co-orbitals. Three of the QS-type Earth co-orbitals (2004 GU$_9$, 2006
FV$_{35}$, 2016 HO$_3$) and  five of the HS-type Earth co-orbitals (2001
GO$_2$, 2002 AA$_{29}$, 2003  YN$_{107}$, 2015 SO$_2$, 2015 YA) show repeated
transitions between QS and HS type orbits over time
\citep{brasser2004transient,de2013resonant}. This repetitive orbit-type
transition theoretically proposed by \citet{namouni1999secular} for the first
time has been discussed in detail in various papers for different asteroids
\citep{connors2002discovery,brasser2004transient,wajer20092002,wajer2010dynamical}.
These asteroids spend most of their co-orbital lifetime in HS co-orbital state.
As a natural consequence of this behaviour, statistically, the HS-type
co-orbital population of those asteroids that show HS-QS transitions can be
expected to appear more frequently than the pure QS-type state
at any given moment during their co-orbital lifetime.

A plausible explanation for the relative difference of abundance of HS versus
TP-type orbits for Earth co-orbitals has been offered by \citet{zhou2019orbital}.
The chaotic motion in the inner Solar system changes the secular frequencies of
the inner planets. This change is known as the frequency drift of the inner
planets. The TP orbits are more sensitive to this drift motion of secular
frequencies than the HS ones. Consequently, regarding the inner planets 
except for Mars, asteroids cannot survive in a Trojan co-orbital state for a long period of
time, but they can if they follow HS-type orbits. However, this
is not the case for Mars Trojans. To date, there have been identified nine
Trojans of Mars, some of which were found in long-term stable orbits
\citep{christou2013orbital,cuk2015yarkovsky,christou2020population}.

The primary sources of Earth co-orbitals are thought to be the various regions
of the main-belt as in the case of non-co-orbital near-Earth asteroids (NEAs)
\citep{morais2002population,galiazzo2014hungaria}.  However, it has been shown
recently by \mbox{\citet{pokorny2019co}} using numerical simulations, that primordial
Venus co-orbitals could exist. It was also showed that some Mars Trojans can be
primordial bodies \citep{connors2005survey,de2013three}.  Similarly, a
primordial co-orbital population may be possible for Earth.  If such primordial
members among Earth co-orbitals are possible, they should be moving along orbits
that are less-sensitive to frequency drift.  Therefore, the primordial
co-orbitals of Earth are more likely to exist following HS-type orbits. This
possibility increases the attractiveness of the study of HS co-orbitals of
Earth.

In the last three years, approximately 35 new asteroids have been found
following orbits whose orbital periods are very close to Earth's (semimajor axes
in the range 0.99 au $< a <$ 1.01 au).
Over the last 35 years, this number has reached 100 and although many of them do
not display co-orbital dynamical behaviour of any kind, 18 have been shown to be
trapped in a 1:1 mean-motion resonance with Earth.
In particular, the low-eccentricity ones among these bodies may be members of
a resonant family defined in \citet{de2013resonant} that is a subgroup of the
proposed near-Earth asteroid belt
\citep{rabinowitz1993evidence,brasser2008asteroids}.

\begin{table*}
 \scriptsize 
  \centering
	\caption{Horseshoe Co-orbitals of Earth (Epoch: JD 2458600.5
	(2019-Apr.-27.0) TDB (J2000.0 ecliptic and equinox). Source: JPL's SSDG
	SBDB)} 
   \label{tab:hscoorbitals}
	\setlength\tabcolsep{1.0pt}
 \begin{tabular}{@{}l r r r r r c c c@{}}
 \toprule
 \textbf{Name} & \multicolumn{1}{c}{\textbf{$a$}} & \multicolumn{1}{c}{\textbf{$e$}} &
	 \multicolumn{1}{c}{\textbf{$i$}}& \multicolumn{1}{c}{\textbf{$\Omega$}}
	 & \multicolumn{1}{c}{\textbf{$\omega$ }} &
	 \textbf{H}&\textbf{MOID}&\textbf{Condition}\\

	 &  \multicolumn{1}{c}{\textbf{(au)}}   &     &
	 \multicolumn{1}{c}{\textbf{(${\degr}$)}} &
	 \multicolumn{1}{c}{\textbf{(${\degr}$)}}
	 &\multicolumn{1}{c}{\textbf{(${\degr}$)}} &
	 \multicolumn{1}{c}{\textbf{(mag)}}  &  \multicolumn{1}{c}{ \textbf{(au)}}   & \multicolumn{1}{c}{\textbf{Code}}\\
               \midrule
(3753) Cruithne & $0.99769415384$ $\pm$ $1.95\times10^{-9}$ & $0.514809615$  $\pm$ $4.86\times10^{-7}$ & $19.80565264$ $\pm$ $1.13\times10^{-5}$ & $126.2272059$ $\pm$ $8.78\times10^{-5}$& $ 43.8400709$ $\pm$ $8.76\times10^{-5}$ &
	 $15.6$ & $0.07120$ & $0$ \\ 
(54509) YORP &$1.005989555595$ $\pm$ $2.60\times10^{-10}$ & $0.229967663$ $\pm$ $1.42\times10^{-7}$ &$ 1.59947076$ $\pm$ $5.19\times10^{-6}$ & $278.2339973$ $\pm$ $6.70\times10^{-5}$ & $278.9371181$ $\pm$ $6.43\times10^{-5}$ & $22.7$ & $0.00276$ & $0$ \\ 
2001 GO$_{2}$ & $1.0066417$ $\pm$ $3.07\times10^{-5}$ & $0.168236$ $\pm$ $6.09\times10^{-4}$ &$4.6250$ $\pm$ $1.93\times10^{-2}$ & $193.54017$ $\pm$ $1.25\times10^{-3}$ & $265.4661$ $\pm$ $2.28\times10^{-2}$ & $24.3$ & $0.00392$ & $7$ \\
2002 AA$_{29}$ & $0.9925404414$ $\pm$ $2.41\times10^{-8}$& $0.013021635$ $\pm$ $2.00\times10^{-7}$ & $10.7482426$ $\pm$ $5.53\times10^{-5}$& $106.3655230$ $\pm$ $2.19\times10^{-5}$& $101.901700$ $\pm$ $4.35\times10^{-4}$ & $24.1$ & $0.01184$ & $0$ \\
2003 YN$_{107}$	& $0.9886958239$ $\pm$ $3.59\times10^{-8}$ & $0.013948370$ $\pm$ $2.11\times10^{-7}$ &$4.3212546 $ $\pm$ $2.63\times10^{-5}$ & $264.4003970$ $\pm$ $6.77\times10^{-5}$ & $ 87.687639$ $\pm$ $1.10\times10^{-4}$& $26.5$ & $0.00504$ & $1$ \\
2006 JY$_{26}$ & $1.0102796$ $\pm$ $1.02\times10^{-5}$ & $0.0830891$ $\pm$ $1.79\times10^{-5}$ & $ 1.438828$ $\pm$ $1.56\times10^{-4}$ & $ 43.46553$ $\pm$ $7.92\times10^{-3}$& $273.6480$ $\pm$ $1.79\times10^{-2}$& $28.4$ & $0.00011$ & $3$ \\
2010 SO$_{16}$ & $1.00316220453$ $\pm$ $8.89\times10^{-9}$ & $0.0754402873$ $\pm$ $9.89\times10^{-8}$ & $14.5183107$ $\pm$ $2.47\times10^{-5}$& $ 40.3844703$ $\pm$ $2.96\times10^{-5}$ & $109.0117369$ $\pm$ $6.34\times10^{-5}$& $20.5$ & $0.02968$ & $0$ \\
2013 BS$_{45}$ & $0.9917335720$ $\pm$ $1.44\times10^{-8}$  & $0.083749794$ $\pm$ $3.30\times10^{-7}$  & $0.77255459$ $\pm$ $3.74\times10^{-6}$ & $ 83.364994$ $\pm$ $4.33\times10^{-4}$& $150.694328$ $\pm$ $4.58\times10^{-4}$& $25.9$ & $0.01147$ & $0$ \\ 
2015 SO$_{2}$ & $0.9956965867$ $\pm$ $3.14\times10^{-8}$ & $0.108714689$ $\pm$ $7.64\times10^{-7}$ &$9.1639445 $ $\pm$ $7.62\times10^{-5}$ & $182.7728701$ $\pm$ $2.19\times10^{-5}$& $291.4372326$ $\pm$ $4.43\times10^{-5}$ & $23.9$ & $0.01889$ & $1$ \\ 
2015 XX$_{169}$	& $1.003154825$ $\pm$ $1.97\times10^{-8}$ & $0.18482325$ $\pm$ $1.51\times10^{-6}$ &$7.6016506$ $\pm$ $4.26\times10^{-5}$ & $256.47577755$ $\pm$ $7.38\times10^{-6}$& $282.890358$ $\pm$ $1.05\times10^{-4}$ & $27.4$ & $0.01500$ & $0$ \\ 
2015 YA & $0.99581349$ $\pm$ $9.08\times10^{-6}$ & $0.279672$ $\pm$ $2.22\times10^{-4}$ &$ 1.619157$ $\pm$ $8.38\times10^{-4}$ & $255.01138$ $\pm$ $3.31\times10^{-3}$& $ 83.6267$ $\pm$ $1.62\times10^{-2}$& $27.4$ & $0.00355$ & $6$ \\ 
2015 YQ$_{1}$ & $1.0039842$ $\pm$ $1.58\times10^{-5}$& $0.403656$ $\pm$ $2.14\times10^{-4}$&$2.48388$ $\pm$ $1.85\times10^{-3}$ & $88.860853$ $\pm$ $3.22\times10^{-4}$& $111.8716$ $\pm$ $1.71\times10^{-2}$ & $28.0$ & $0.00061$ & $6$ \\
          \bottomrule\\
   \end{tabular}
\end{table*}

In this work, we have studied near-Earth asteroids moving along orbits with
semimajor axes in the range 0.983 au $< a <$ 1.017 au to find among them those
that are candidates to be trapped in 1:1 mean-motion resonance with Earth.
Six objects, namely 2016 CO$_{246}$, 2017 SL$_{16}$, 2017 XQ$_{60}$, 2018
PN$_{22}$, 2018 AN$_2$ and 2018 XW${_2}$ moving along HS-type orbits have been
identified.  The orbits of asteroids 2016 CO$_{246}$, 2017 SL$_{16}$, 2017
XQ$_{60}$ and 2018 PN$_{22}$ with relatively good orbital solutions have been
studied in detail. As a result of the dynamical analyses of their orbital
evolutions, these asteroids can be classified as HS co-orbitals of Earth.
Section 2 includes a short review about the already known HS co-orbitals of
Earth.  In Sect. 3, the initial conditions, and numerical methods used are
outlined.  Section 4 presents and discuss the results. Section 5 summarizes the
conclusions.

\section{A short review of Horseshoe Co-orbitals of Earth}
\label{ShortRew}

If a small body is in a 1:1 mean motion resonance with a planet, and librates in
an orbit that covers the Lagrangian points of the planet L4, L5, and L3, the
regular horseshoe-type co-orbital state is appropriate to describe such an
orbit. Even if horseshoe orbits are not considered to be long-term stable, their
probabilities of survival depend on the mass of the host planet. These
probabilities are higher for the cases of hosts with smaller masses like those
of Earth and Venus 
\citep{dermott1981dynamicsa,dermott1981dynamicsb,murray1999solar,cuk2012long,
de2013crantor}.

The HS co-orbitals are interesting not only due to the shape of their
orbital paths but also because they have peculiar dynamical properties.  For
instance, within the framework of the co-orbital motion, the maximum libration
range of the semimajor axis ({\it a}) is seen in the HS-type orbits.
Moreover, there may be even a ten-times difference or more between the points of
closest approach to the planet and farthest distance from it in the case of HS
orbital paths.  Regarding this latter peculiar dynamical property, the minimum
and maximum distances between (419624) 2010 SO$_{16}$ and Earth in one HS cycle
are 0.204 au and 2.05 au, respectively.

Osculating orbital elements, absolute magnitude (H), minimum orbit intersection
distances for Earth (MOID), and condition codes (a measure of orbit
uncertainty in JPL Horizon database, the uncertainty increases from 0 to 9) of
the documented HS co-orbitals of Earth have been listed in Table~\ref{tab:hscoorbitals}.
The MOID values of all but one, (3753) Cruithne, of the known
HS co-orbitals verify the necessary distance criterion (<0.05 au) to be 
PHA (Potentially Hazardous Asteroid). However, they are not considered PHA due
to their absolute magnitude, except (419624) 2010 SO$_{16}$.  By definition, in
order for an asteroid to be considered PHA, its absolute magnitude must be  22.0
or brighter. The fact that they are not potentially dangerous on a global scale
does not imply that they will have no effect on Earth in the event of an impact
or approach well within the Hill radius of Earth. However, small objects can not
be observed as often as large bodies \citep{de2016trio}.

\begin{figure}
	\includegraphics[width=\textwidth,height=0.6\textwidth,keepaspectratio]{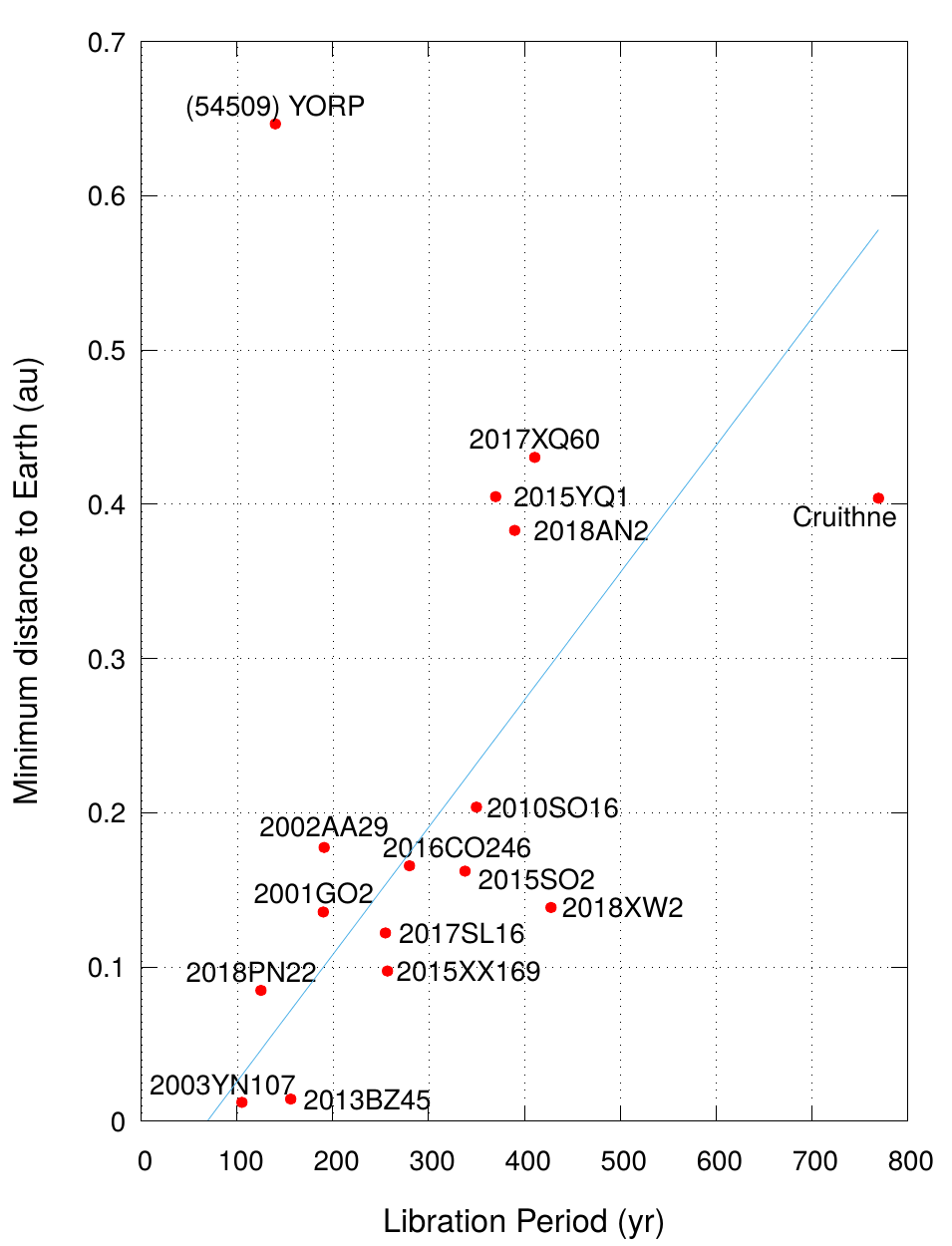}
	\caption{Libration period versus minimum distance to Earth for HS
	co-orbitals of Earth. The blue line corresponds to a linear fit for
	those whose eccentricity value is less than $0.2$ ((3753) Cruithne,
	(54509) YORP, 2015 YQ$_{1}$, 2017 XQ$_{60}$, and 2018 XW$_{2}$ are excluded).}
    \label{fig:lib_periods}
\end{figure}

It was shown in \citet{hollabaugh1973earth} for hypothetical Earth
co-orbital asteroids that the libration periods of HS type orbits depend on how
close the asteroid approaches Earth. The average inclination is 6${\degr}$ and
initial eccentricity is 0 in their numerical simulations.
This relation between the minimum distances of the asteroids to Earth and their
libration periods for both current Earth HSs (Table~\ref{tab:hscoorbitals}) and
our HS candidates (Table~\ref{table:testasteroids}) is shown in
Figure~\ref{fig:lib_periods}.
The blue line in Figure~\ref{fig:lib_periods} which corresponds to a
linear fit for those whose eccentricity value is less than 0.2 has a quite
similar slope for the same relation in Fig.~2 of \citet{hollabaugh1973earth}.

Another peculiar dynamical feature of the HS co-orbitals of Earth is that
the members of this subgroup of minor bodies are periodically switching between
Apollo and Aten dynamical classes. At the time of this writing, six of the
known HS co-orbitals of Earth are classified as members of the Apollo class,
being the remaining HS co-orbitals members of the Aten class.

The criteria to classify an object as either member of Apollo or Aten dynamical
class are based on the values of their perihelion and aphelion distances, and
semimajor axes as compared to those of Earth. If $q$ (perihelion distance)
of the asteroid is less than Earth's aphelion distance ($q <$ 1.017 au) but
$a$ is greater than that of Earth ($>$ 1 au), it is classified as Apollo
asteroid. On the other hand, if $Q$ (aphelion distance) of the asteroid is
greater than Earth's perihelion distance ($Q >$ 0.983 au) but $a$ is less
than that of Earth ($<$ 1 au), it is classified as an Aten asteroid.

This clearly indicates that, for all objects in the range $0.983 $ au $<a<$
$1.017$ au, the necessary perihelion and aphelion limits for being an Apollo or
Aten respectively are ensured.  
In addition, it should be pointed out here that the current values of $a$ and
libration lengths for all objects known to be Earth co-orbitals are between
0.983 au and 1.017 au. This implies that the classification of co-orbitals
moving along HS-type orbits as members of the Apollo or Aten class depends on
whether their values of $a$ are currently greater than or less than 1 au.
In fact and due to the oscillation of $a$ as a consequence of being trapped in a
1:1 mean-motion resonance with Earth, they are switching periodically between
the Apollo and Aten classes. For instance, 2016 CO$_{246}$, one of the objects
analysed in this work, is in the Apollo class for Epoch 2017-Jan-13.0 and in the
Aten class for Epoch 2018-Sep-10.

It is relatively frequent for some asteroids to switch between Apollo class and
Aten class membership. However, this is usually caused by the chaotic orbital
movements of asteroids.  The situation mentioned here is the result of the
almost-periodic motion of the objects as a consequence of being trapped in a 1:1
mean-motion resonance with Earth.

According to the classical definition of co-orbital motion
\citep[see][]{murray1999solar}, two objects that are co-orbital candidates have
to share the value of the semimajor axis as well as to have a relative mean
longitude that librates around certain particular values.
The mean longitude of an object is $\lambda$ = $M$ + $\Omega$ + $\omega$, where 
$M$ is the mean anomaly, $\Omega$ is the longitude of the ascending node and 
$\omega$ is the argument of perihelion. The relative mean longitude is   
defined as the difference between the mean longitude of the host body and the
mean longitude of the object. 						     
If it librates around 0${\degr}$, the object is in the quasi-satellite co-orbital state, if it
librates around 60${\degr}$, the object is called an L4 Trojan, when it 
librates around 300${\degr}$ (or -60${\degr}$), it is an L5 Trojan, whereas 
if the libration amplitude is larger than 180${\degr}$ is called a horseshoe
\citep{dermott1981dynamicsa,dermott1981dynamicsb,murray1999solar,morais2002population,connors2002discovery}.
Although objects in co-orbital motion share the semimajor axis and their mean
longitudes relative to the host body, Earth in this case, oscillate around
certain values, i.e. either 0${\degr}$, 60${\degr}$, -60${\degr}$, or
180${\degr}$, their eccentricity and inclination may be very different.
Consequently, the asteroids (3753) Cruithne, (54509) YORP (2000 PH$_{5}$) listed
here (Table~\ref{tab:hscoorbitals}) have very high orbital eccentricities and
are thus moving along orbits that are different from that of Earth.	

In particular, these bodies can be considered as HS-type co-orbitals of Earth,
being in a 1:1 mean-motion resonance with Earth and having as shape of their
orbit projected onto the ecliptic plane, from the point of view of a frame of
reference that co-rotates with Earth, one that resembles that of an actual
horseshoe. This first group of objects is called unusual HS co-orbitals. 

The orbits of 2002 AA$_{29}$, 2003 YN$_{107}$, (419624) 2010 SO$_{16}$, (454094)
2013 BS$_{45}$, and 2015 SO$_{2}$ are well suited for these objects to be
classified as HS-type co-orbitals of Earth in the classical sense.
Their eccentricities are quite small. The maximum $e$ value in this group is
$0.1087152$ which belongs to asteroid 2015 SO$_{2}$. 
Their relative mean longitude librates with an amplitude larger than
180${\degr}$, and they are in 1:1 mean-motion resonance with Earth.
However, they do not pass between Earth and the Sun in the frame of
reference that co-rotates with Earth.
Orbits of these objects can be termed {\it nearly-symmetric HS}.

There exists a third subgroup of Earth co-orbitals considered to be HS. The
orbital eccentricities of these bodies are greater than that of the
nearly-symmetric HS and smaller than that of the unusual HSs. In contrast to
non-symmetrical HS orbits, the relative mean longitudes of these bodies
reach $0{\degr}$. This implies that the "horns" in the relative orbit shape of the
objects are not positioned to hold Earth in between. Therefore, the object
crosses twice the minimum distance with Earth in a HS cycle. 2001
GO$_{2}$, 2015 XX$_{169}$ and 2006 JY$_{26}$ are the members of this group. For
the orbit of these bodies, as mentioned in \citet{de2016trio}, the {\it
asymmetrical HS} expression seems to fit.

There are two objects which we have not counted yet in any of these three
classes. These objects have been classified as HS in the literature, they are
2015 YA and 2015 YQ$_{1}$.

Analyses and evaluations of the dynamical evolution of asteroid 2015 YA have
been presented by \citet{de2016trio}. The orbital solutions of the 2015 YA were
relatively poor at that time. Thus, some statistical approaches were used to
assess the dynamical state of the asteroid according to the then nominal
parameters.  In the above mentioned work, it is stated that the object follows
an asymmetrical horseshoe path viewed in a frame of reference that co-rotates
with Earth and projected onto the ecliptic plane. Even if the orbital solution
of 2015 YA is still not so well defined, uncertainties seem to have diminished a
little over time.  At least, the new nominal orbital parameters produce somewhat
different results than the ones mentioned in that paper.

According to the orbital solution that corresponds to the current nominal
orbital parameters, 2015 YA stays near Earth's mean longitude for almost
$225$ yr. This implies that its orbit follows an asymmetrical quasi-satellite
trajectory with respect to Earth. After $225$ yr of dynamical evolution, its
orbital path starts to show asymmetrical HS features.

Additionally, a similar situation is observed for 2015 YQ$_{1}$. The orbit is
defined as asymmetrical HS in \citet{de2016trio}. While the 2015 YQ$_{1}$ seems
to be an asymmetrical HS according to the calculations made for its
current nominal orbital  elements, the "horns" of the HS are located behind
the L4 point. The object should be classified as {\it unusual HS} due to its
high $e$ value and irregular $a$ graph.

\section{Numerical Methods and Initial Conditions}

\begin{table*}
	\scriptsize
 \centering
	\caption{Orbital elements of HS co-orbital candidates (Epoch:JD
	2458600.5 (2019-Apr.-27.0) TDB (J2000.0 ecliptic and equinox). Source: JPL's
	SSDG SBDB)}
    \label{table:testasteroids}
	\setlength\tabcolsep{2.5pt}
 \begin{tabular}{@{}l r r r r r c c c@{}}
  \toprule
 \textbf{Name} & \multicolumn{1}{c}{\textbf{$a$}} & \multicolumn{1}{c}{\textbf{$e$}} &
	 \multicolumn{1}{c}{\textbf{$i$}}& \multicolumn{1}{c}{\textbf{$\Omega$}}
	 & \multicolumn{1}{c}{\textbf{$\omega$ }} &
	 \textbf{H}&\textbf{MOID}&\textbf{Condition}\\

	 &  \multicolumn{1}{c}{\textbf{(au)}}   &     &
	 \multicolumn{1}{c}{\textbf{(${\degr}$)}} &
	 \multicolumn{1}{c}{\textbf{(${\degr}$)}}
	 &\multicolumn{1}{c}{\textbf{(${\degr}$)}} &
	 \multicolumn{1}{c}{\textbf{(mag)}}  &  \multicolumn{1}{c}{ \textbf{(au)}}   & \multicolumn{1}{c}{\textbf{Code}}\\
               \midrule
2016 CO$_{246}$& $0.99882330264$ $\pm$ $4.05\times10^{-9}$& $0.1256830$ $\pm$ $4.00\times10^{-7}$& $6.3271019$ $\pm$ $1.76\times10^{-5}$& $136.6355005$ $\pm$ $1.06\times10^{-5}$ & $119.4946322$ $\pm$ $2.61\times10^{-5}$& $25.8$& $.03855$&  $0$\\
2017 XQ$_{60}$ &$1.00032944453$ $\pm$ $6.52\times10^{-9}$ & $0.21413061$ $\pm$ $2.61\times10^{-6}$& $27.199615$ $\pm$ $2.55\times10^{-4}$& $269.20693549$ $\pm$ $9.26\times10^{-6}$& $283.554789$ $\pm$ $1.64\times10^{-4}$& $24.4$& $.01931$&  $0$\\
2017 SL$_{16}$ &$1.0003248237$ $\pm$ $1.08\times10^{-8}$ & $0.15316581$ $\pm$ $1.05\times10^{-6}$& $8.6827950$ $\pm$ $5.48\times10^{-5}$& $182.5808557$ $\pm$ $1.03\times10^{-5}$& $70.4223009$ $\pm$ $4.86\times10^{-5}$  & $25.8$ & $.01765$&  $0$\\
2018 PN$_{22}$ &$0.997175533$ $\pm$ $4.93\times10^{-7}$& $0.0392063$ $\pm$ $1.08\times10^{-5}$ &$4.38465$ $\pm$ $1.74\times10^{-3}$& $317.07727$ $\pm$ $1.09\times10^{-3}$ & $219.1738$ $\pm$ $1.32\times10^{-2}$ & $27.5$ & $.01156$&  $3$\\
2018 XW$_{2}$  &$0.998466$ $\pm$ $3.67\times10^{-4}$& $0.301953$ $\pm$ $5.23\times10^{-4}$& $19.7295$ $\pm$ $3.98\times10^{-2}$ & $78.17736$ $\pm$ $2.22\times10^{-3}$& $250.02149$ $\pm$ $3.06\times10^{-2}$& $25.5$ & $.01952$&  $7$\\
2018 AN$_{2}$  & $1.001448$ $\pm$ $3.11\times10^{-4}$& $0.154277$ $\pm$ $2.54\times10^{-4}$& $22.0770$ $\pm$ $4.65\times10^{-2}$ & $290.12682$ $\pm$ $4.73\times10^{-3}$ &$284.7790$ $\pm$ $8.38\times10^{-2}$ & $24.8$& $.03179$&  $7$\\
          \bottomrule\\
   \end{tabular}
\end{table*}

It cannot be claimed that a minor body is in co-orbital motion with a host
planet simply because its orbital elements have certain particular values.
The dynamical evolution of some orbital parameters must be studied by means of
forward and backward integrations over a reasonably long amount of time. 
In particular, the time evolution of $a$ and $\lambda_r$ (mean longitude
relative to Earth) provides information about whether the orbit is
co-orbital. The orbit is integrated both forward and backward in time;
$\lambda_r$ is followed. If the asteroid is in a co-orbital state, $\lambda_r$
librates around $\pm 60{\degr}$ (or 300${\degr}$), $\pm 180{\degr}$, or $0{\degr} $, and it is
classified as tadpole, horseshoe, or quasi-satellite, respectively
\citep{de2016horseshoe}.

Small uncertainties in the orbital elements may cause relatively large
differences in the results after a medium or long-term integration due to the
chaotic nature of the dynamical evolution.  Thus, it is necessary to investigate
the effects of small uncertainties on the dynamical evolution of the orbital
elements using clones within the framework of a statistical approach.  As a
result of these statistical analyses, a more reliable result is obtained about
the shape and evolution of the orbit. In this way, it is possible to gain a
better insight into the orbital stability in the close vicinity of the nominal orbit of
the co-orbital candidate.

In the calculation of the probability distribution, a clone must complete
at least one HS libration period in backward or forward simulations starting from
its current position to be assumed that it is moving along a HS type orbit. 
The probability of one body being in a co-orbital state is obtained by dividing
the number of orbits in this state by the total number of clones. The total
probability is therefore only valid for one libration period.

In this work, the MCCM (Monte Carlo using Covariance Matrix)
\citep{bordovitsyna2001new,avdyushev2007regions,de2012dynamical} method is
used to produce clone orbits. Through this approach, the effects of the
uncertainties of the orbital parameters on other parameters are included in the
generated clones.

A model Solar system including the eight known planets, the dwarf planet (1)
Ceres, the main-belt asteroids (2) Pallas, (4) Vesta, (10) Hygiea, (31)
Euphrosyne and the Moon has been used for the numerical integrations performed
with the REBOUND $N$-body integration package \citep{rein2012rebound}. 
This package makes use of a 15th order Gau{\ss}-Radau quadrature
integration scheme (IAS15 integrator) \citep{rein2014ias15}.  
The initial conditions for all large and dwarf planets, minor bodies and the
Moon have been obtained using the Jet Propulsion Laboratory's Solar System
Dynamics Group Small-Body Database (JPL's SSDG SBDB,
\citealp{giorgini2001orbit,giorgini2011proceedings,giorgini2015iau}) and JPL's
Horizons ephemeris system \citep{giorgini1996jpl,Standish1998,giorgini1999line}
for the epoch JD 2458600.5 (2018-Apr-27.0) TDB (Barycentric Dynamical Time).

Some tests were made for known HS co-orbitals of Earth with both MERCURY 6
package \citep{chambers1999hybrid} with Bulirsch-Stoer integrator and REBOUND
package with IAS15 integrator to check the reliability of the results. We got
almost the same results for both packages for a few thousand years. However,
relative energy error for IAS15 integrator was in machine precision level.
Besides, tests were performed using REBOUND package with IAS15 integrator for
known HS co-orbitals of Earth and the results were compared with those in the
literature. Notable different results were obtained for only two bodies (2015 YA
and 2015 YQ$_1$) and they were mentioned in Section 2.

Numerical integration tests showed that the inclusion of dwarf planets and the
main-belt asteroids has no significant effect on the results of the dynamical
evolution of the minor bodies studied here within the time-scale of dynamical
evolution studied in this research. 

The upper and lower limits for the semimajor axis of an asteroid in a 1:1 mean
motion resonance with Earth are given as $0.99$ au and $1.01$ au
\citep{morais2002population}. However, our preliminary studies have shown that
the semimajor axes librations for asteroids 2006 JY$_{26}$ and 2018 PN$_{22}$
force these limits. Thus, in this study the surveyed part of the orbital
parameter space corresponds to a semimajor axis range of 0.983 au $< a <$ 1.017
au, slightly larger than the theoretical limit. 
At the time of this writing (2019 April 1) there are 239 minor bodies 
in this semimajor-axis range (source: JPL Small-Body Database Search Engine). 
We have studied the dynamical evolution of these near-Earth objects and found
that the evolution of the nominal orbital parameters of asteroids 2016
CO$_{246}$, 2017 SL$_{16}$, 2017 XQ$_{60}$, 2018 PN$_{22}$, 2018 AN$_{2}$ and
2018 XW$_{2}$ is consistent with them being trapped in a 1:1 mean-motion
resonance with Earth and showing dynamical features characteristic of the
HS-type dynamical state.
Table~\ref{table:testasteroids} shows the osculating orbital parameters, absolute
magnitudes, and condition codes of the HS co-orbital candidates for
Earth. 
At the moment, the orbital solutions for the asteroids 2018 AN$_{2}$ and
2018 XW$_{2}$ are still poor.
This implies that their condition codes (uncertainty parameters) are
currently high (both 7). If they are not observed again in the next future and
at the appropriate positions, their  condition codes will stay at the same level
for a long time like in the case of 2001 GO$_{2}$ \citep{brasser2004transient}.

\section{Results}
Here, nominal and clone orbits of asteroids 2016 CO$_{246}$, 2017 SL$_{16}$, 
2017 XQ$_{60}$, and 2018 PN$_{22}$ with low orbital uncertainties have
been analysed by numerically integrating their orbits both backward and forward
in time. To clarify the possible scattering in orbital parameters that will
occur over time in the integration process of clone orbits, $1000$ different
clone orbits for forward and backward integrations have been produced and used.

\begin{figure*}
    \begin{subfigure}[b]{1.0\linewidth}
	\includegraphics[width=\linewidth,height=0.29\linewidth]{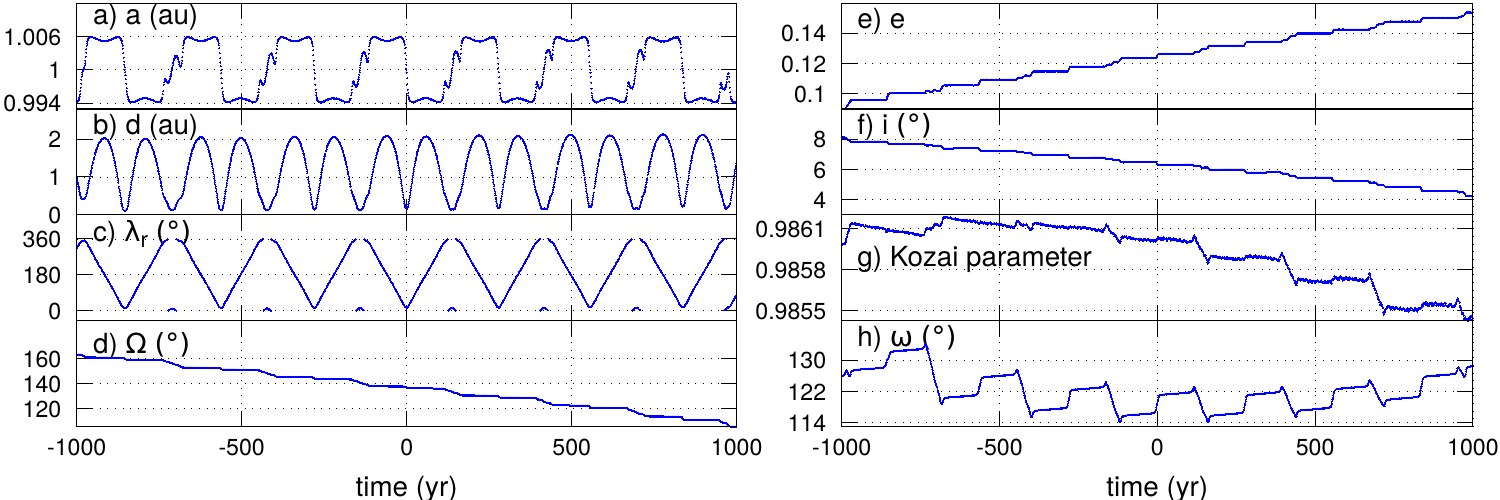}
	    \caption{ 2016 CO$_{246}$ }
	     \label{fig2:2016CO246}
     \end{subfigure}
     \begin{subfigure}[b]{1.0\linewidth}
	\includegraphics[width=\linewidth,height=0.29\linewidth]{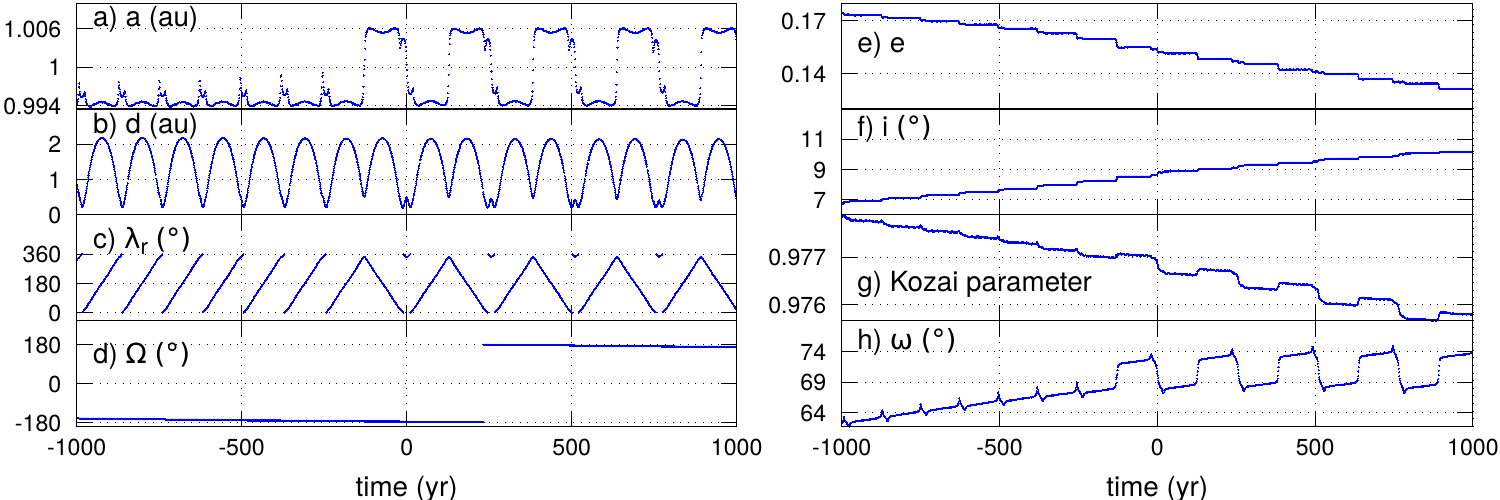}
		\caption{2017 SL$_{16}$}
		 \label{fig2:2017SL16}
     \end{subfigure}
     \begin{subfigure}[b]{1.0\linewidth}
	\includegraphics[width=\linewidth,height=0.29\linewidth]{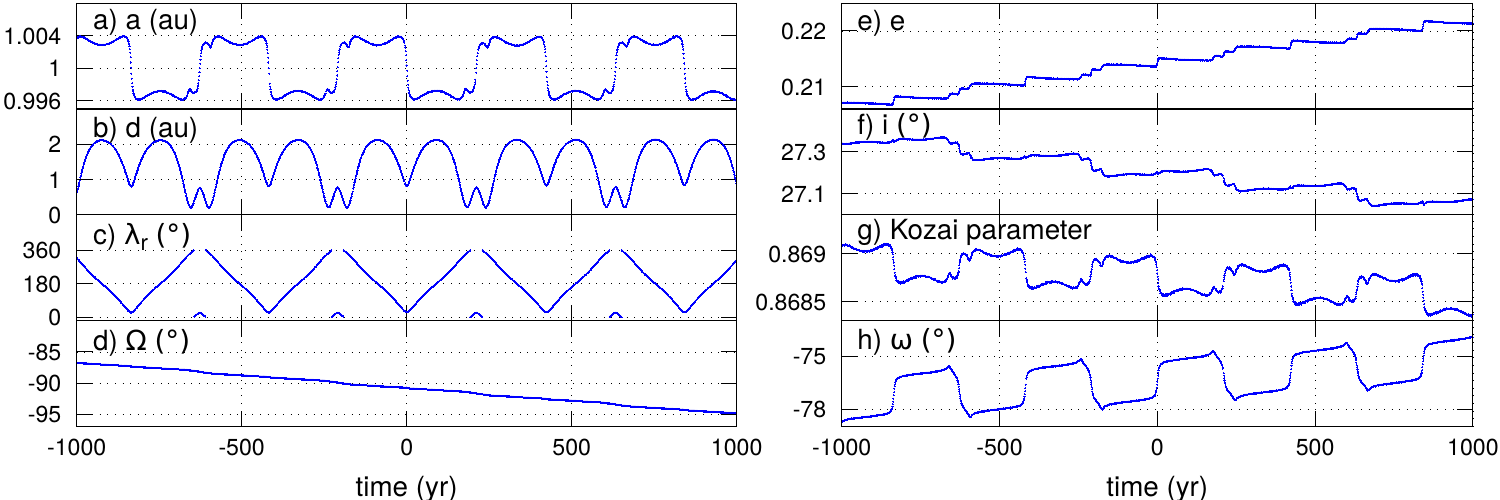}
		\caption{2017 XQ$_{60}$}
		 \label{fig2:2017XQ60}
     \end{subfigure}
     \begin{subfigure}[b]{1.0\linewidth}
	\includegraphics[width=\linewidth,height=0.29\linewidth]{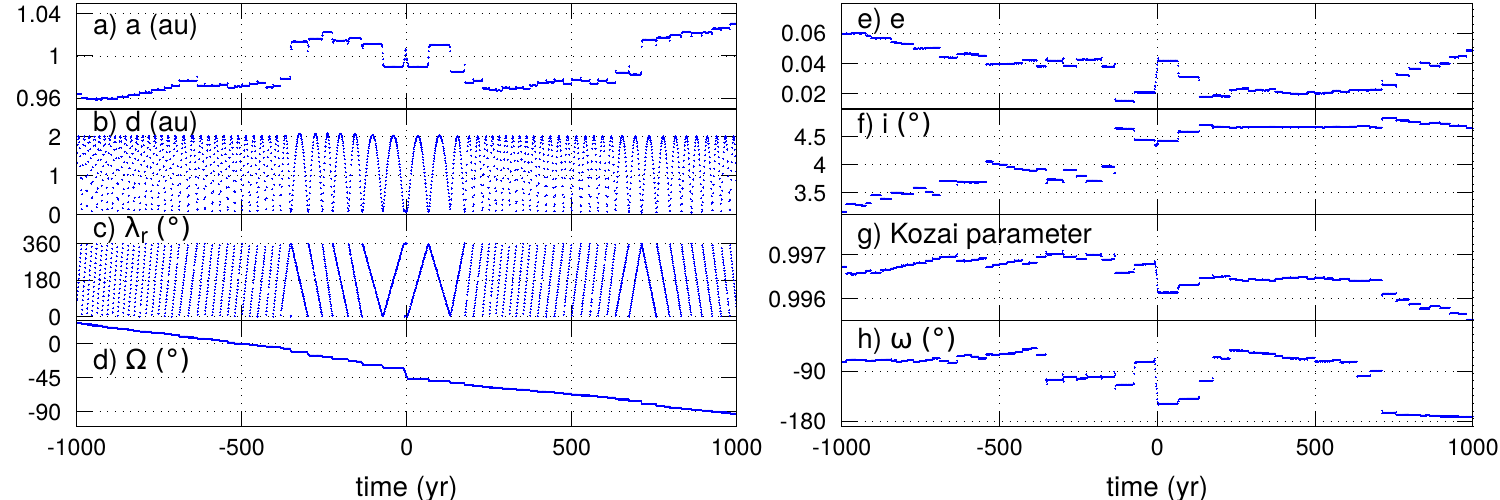}
	\caption{2018 PN$_{22}$}
	     \label{fig2:2018PN22}
     \end{subfigure}
	    \caption{
		    Short-term dynamical evolution of some parameters for the nominal
		    orbits of 2016 CO$_{246}$, 2017 SL$_{16}$, 2017 XQ$_{60}$,
		    and 2018 PN$_{22}$. The five basic orbital parameters ($a$,
		    $e$, $i$, $\Omega$, $\omega$) are plotted together with the
		    distance from Earth ($d$), the relative mean longitude
		    ($\lambda_r$) and the Kozai parameter $\sqrt{1-e^2}\cos i$.}

      \label{fig:201620172018}
\end{figure*}

Figure~\ref{fig:201620172018} shows the short-term dynamical evolution of the
relevant orbital parameters $a$, $e$, $i$, $\Omega$, $\omega$, $d$ (Earth
distance in au), $\lambda_r$, and the Kozai parameter ($\sqrt{1-e^2}\cos i$)
for the nominal orbits of 2016 CO$_{246}$, 2017 SL$_{16}$, 2017 XQ$_{60}$, 
and 2018 PN$_{22}$ for a time interval of integration that goes from 1000 yr
into the past to 1000 yr into the future.

Figures~\ref{fig:2016CO246aeil_mean}, \ref{fig:2017SL16aeil_mean},
\ref{fig:2017XQ60aeil_mean}, and \ref{fig:2018PN22aeil_mean} show the dynamical
evolution of the orbital parameters for 2016 CO$_{246}$, 2017 SL$_{16}$, 2017
XQ$_{60}$, and 2018 PN$_{22}$ with the averages of their clones' orbital
parameters for $\mp 10000$ yr. The clones were produced according to the
MCCM method. 1000 different clones have been used for forward and backward
integrations. In addition to the relevant orbital parameters, we have also
studied the dynamical evolution of the Kozai parameter.

In addition to these, some long-term integrations were performed for 2016
CO$_{246}$, 2017 SL$_{16}$ and 2017 XQ$_{60}$ for 1 Myr using 10 clon
orbits for each body.
However, small differences between clone orbits started to increase in a
few kyr of dynamical evolution. Their dynamical evolutions are chaotic for
long term integrations. As a result, semimajor axes of all the clone orbits of
2016 CO$_{246}$ stayed in the range of 0.992 au to 1.008 au for about 80 kyr. 
For 2017 SL$_{16}$ the range is 0.990 au to 1.007 au for a duration of 18 kyr, 
and for 2017 XQ$_{60}$ the range is 0.994 au to 1.007 au for a duration of 19
kyr.

After that, they began to follow different trajectories. Besides, it should be
considered that these are very small bodies and are expected to be more affected
by non-gravitational effects in long-term integrations. Therefore, very
long-term integrations should also be made using a large number of clones
including non-gravitational effects to obtain reasonable results.

\subsection{Asteroid 2016 CO$_{246}$}
Asteroid 2016 CO$_{246}$ was discovered by the Pan-STARRS survey 
\citep{kaiser2004pana,kaiser2004panb} on 2016 February 11. The orbit
determination of this minor body is based on 42 observations for a data-arc span
of 1088 days. The condition code is 0, which indicates that the uncertainty in
the calculated orbit is rather small.  Asteroid 2016 CO$_{246}$ is relatively
small with an absolute magnitude of 25.8 mag, which suggests a diameter in
the range 12-92 m for an assumed albedo in the range 0.60-0.01. The MOID value
for Earth is $0.03752$ au. However, due to its absolute magnitude, which is less
than $22$, the object is not in the PHA list.

2016 CO$_{246}$ is classified as an Aten in ESA's NEO page and as an Apollo in
the JPL Small-Body Database Browser (its semimajor axis is $a$ = 0.99936 au for
Epoch JD 2458371.5 (2018-Sept.-10.0) but $a$ = 1.00075 for Epoch JD 2457766.5
(2017-Jan.-13.0).  Both classifications are correct due to the fact that the
object switches between Apollo and Aten dynamical classes.
It was an Apollo in 2017 but it is a member of the Aten dynamical class
currently, and our calculations indicate that it will return to the Apollo
dynamical class after 138 yr of dynamical evolution.
As we have already mentioned in Section 2, the classification of the
object will periodically alternate between the Apollo and Aten classes as long
as it is in a HS-type orbit.

The semimajor axis short-term dynamical evolution is the typical one of an
HS-type co-orbital state, being its libration period 280 yr.  What is peculiar
and different here is that there is a second oscillation between the minimum and
maximum values of the semimajor axis.  Due to the fact that the gap between the
horns in its nominal HS orbit is not positioned to hold Earth in between.
As 2016 CO$_{246}$ moves along its orbit, the minor body passes twice through
its point of minimum distance to Earth in one HS period.  Moreover, this
situation is seen in the graph showing the distance of the object from Earth.
This corresponds to the definition of {\it asymmetrical HS} as defined in
Section~\ref{ShortRew}.

Furthermore, the $\lambda_r$ graph (Fig.~\ref{fig2:2016CO246}, panel c) shows
that $\lambda_r$ oscillates around $180{\degr}$, as expected for a minor
body that is in HS co-orbital state. However and for the reasons pointed out
above, when the object reaches the Earth-Sun direction in the relative orbital
plane, the angle $\lambda_r$ reaches $0{\degr}$.  When the body moves on to the end
of the horn of its HS path, $\lambda_r$ reaches about $5{\degr}$. Thereafter,
it returns to $0{\degr}$ and continues moving along its orbit.

\begin{figure*}
	\includegraphics[width=\textwidth]{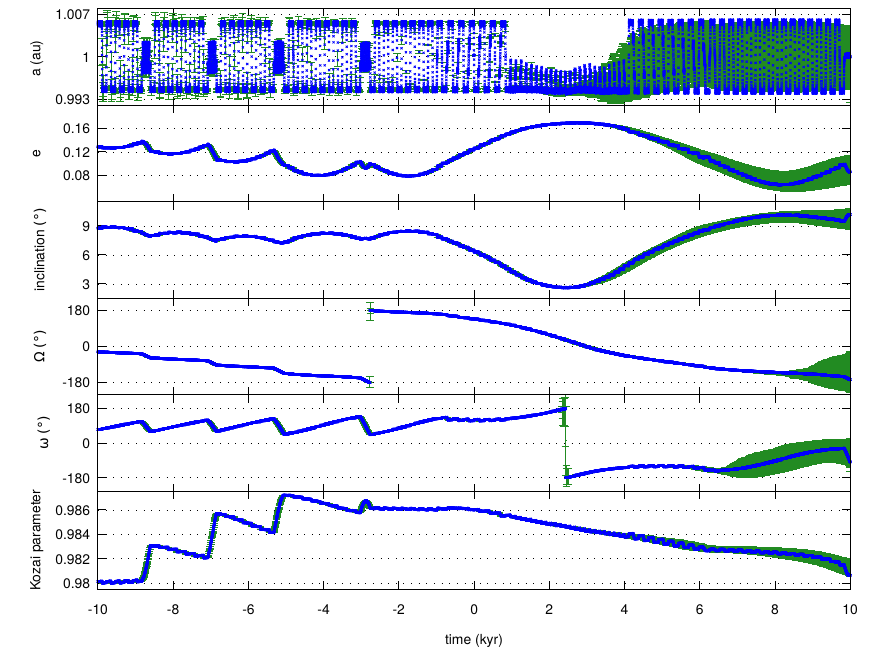}
	\caption{Average of the short-term dynamical
	evolution of $1000$ clone orbits of 2016 CO$_{246}$ ($1000$
	clones for forward and $1000$ clones for backward integrations) 
	and the nominal orbit of 2016 CO$_{246}$.
	The green curves show the average of the parameters with error bars for
	clone orbits; the blue curves show the parameters for the nominal orbit.}
    \label{fig:2016CO246aeil_mean}
\end{figure*}

Fig.~\ref{fig:2016CO246aeil_mean} portrays that the HS dynamical behaviour
is preserved in all the clone orbits. Averages of the clone parameters and the
parameters of the nominal orbit are largely consistent. 
Asteroid 2016 CO$_{246}$ will remain in an HS co-orbital state for about 900 yr.
The object's nominal orbit changes between {\it nearly symmetric HS}, QS,
{\it asymmetrical HS}, and passing object within the time interval -10000 yr- 10000 yr.
Even in the $a$ graph (in Fig.~\ref{fig:2016CO246aeil_mean}), where the largest
difference between the average orbit of the clones and the nominal orbit is
observed, the consistency for the time interval of -10000 yr- 10000 yr is
preserved. After 2500 yr of dynamical evolution, small differences are
increasing, so the average values differ from those of the nominal orbit. 
An interpretation of this situation leads us to conclude that the dynamical
evolution is chaotic as orbits starting arbitrarily close to each other diverge
in a relatively short time-scale, but the overall dynamics resembles that of
stable chaotic orbits \citep{milani1992example}. The stability of the orbit was
higher in the past and the Lyapunov time was also longer in the past.  If we
interpret orbital stability as the object staying in a certain region as in
\citet{de2016asteroid}, all the clones of the object remain trapped in a 1:1
mean-motion resonance with Earth for 20000 yr.

The values of $e$ and $i$ oscillate periodically, alternating high $e$ and $i$.
This dynamical behaviour resembles the motion of a body subjected to a
Lidov-Kozai resonance \citep{kozai1962secular,lidov1962evolution,ito2019lidov}.
However,
mutual increases and decreases in both $e$ and $i$ should be proportional to
each other to keep the Kozai parameter constant. Furthermore, during the same
period, $\omega$ is expected to oscillate around a certain value. This
oscillation can take place at different values for two different configurations.
If the ratio of perturbed and perturber semimajor axes is close to zero,
$\omega$ is expected to oscillate around $90{\degr}$ or $270{\degr}$
($-90{\degr}$). If the asteroid is in 1:1 mean motion resonance with the
perturber, the libration of $\omega$ occurs at around $0{\degr}$ or $180{\degr}$
\citep{de2016asteroid}.

It is seen in Fig.~\ref{fig:2016CO246aeil_mean} that  $\omega$ has oscillated
around 90${\degr}$ in the time interval -10000 yr- -1000 yr. Within this
period, $e$ and $i$ show a proportional increase/decrease. At the same time, the
nominal orbits and the clones indicate that the orbit of the object has made
transitions between the QS and HS co-orbital states. Additionally,
the periods in which the turning points of  reciprocal decreases and increases
in the $e$ and $i$ values are peaked, correspond to the periods of the QS
type orbit. However, the Kozai parameter is likely to increase due to other
reasons.

It is foreseen and well explained by Namouni (1999) and Christou (2000) that
HS-QS transitions can be observed in classical Lidov-Kozai resonance.
Besides, the backward simulation part of Fig.~\ref{fig:2016CO246aeil_mean}
suggesting Lidov-Kozai resonance is similar to the Fig.~4 of 2016 HO$_3$ in
\citet{de2016asteroid}, Fig.~2 of 2015 SO$_2$ in \citet{de2016horseshoe},  and
Fig.~15 in \citet{namouni1999secular}.  It has been shown in the relevant papers
that 2015 SO$_2$ is exposed to Kozai resonance, but the status of 2016 HO$_3$
fits the horseshoe-retrograde satellite orbit transitions described in
\citet{namouni1999secular}.

\begin{figure}
    \begin{subfigure}[b]{1.0\linewidth}
	\includegraphics[width=\linewidth,height=0.45\linewidth]{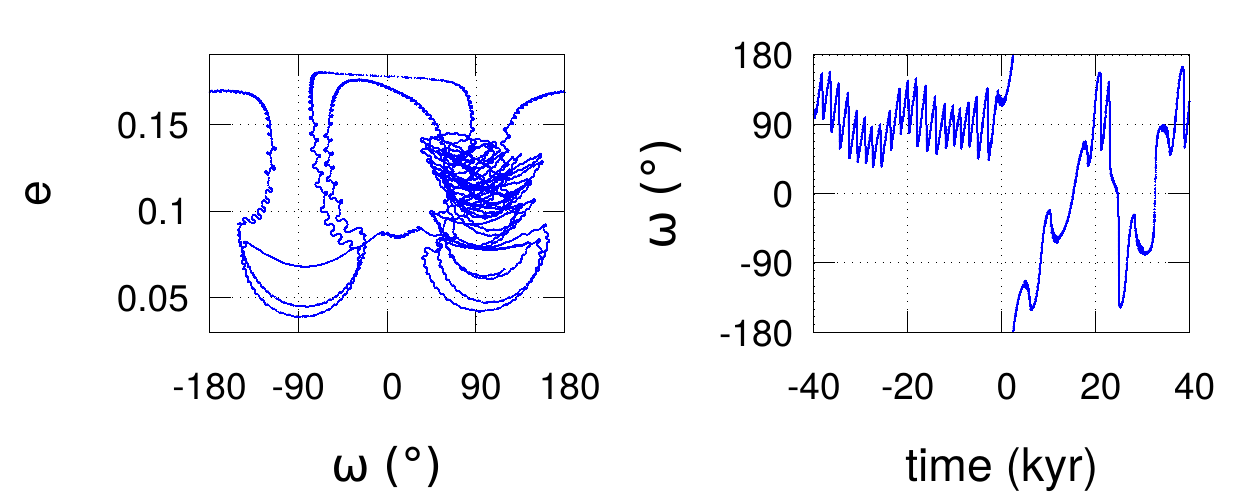}
	    \caption{Full integration for 2016 CO$_{246}$ with all planets and
	    asteroids.}
	     \label{fig:2016CO246_wAll}
     \end{subfigure}
     \begin{subfigure}[b]{1.0\linewidth}
	\includegraphics[width=\linewidth,height=0.45\linewidth]{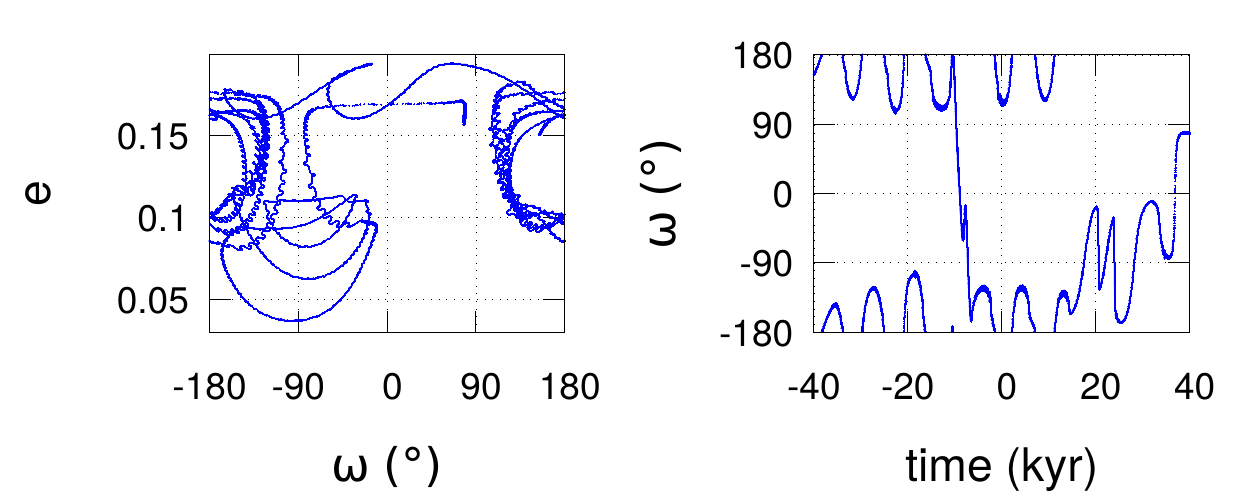}
		\caption{Same as (a) but when Jupiter is removed from the calculations.}
		 \label{fig:2016CO246_noJ}
     \end{subfigure}
     \begin{subfigure}[b]{1.0\linewidth}
	\includegraphics[width=\linewidth,height=0.45\linewidth]{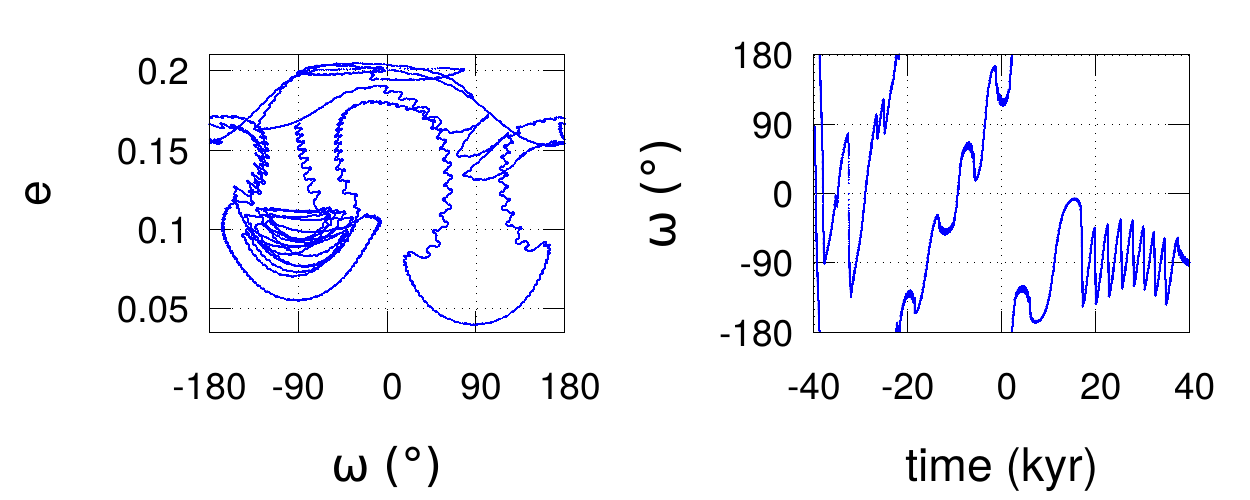}
	     \caption{Same as (a) but when Venus is removed from the calculations.}
	     \label{fig:2016CO246_noV}
     \end{subfigure}
     \begin{subfigure}[b]{1.0\linewidth}
	\includegraphics[width=\linewidth,height=0.45\linewidth]{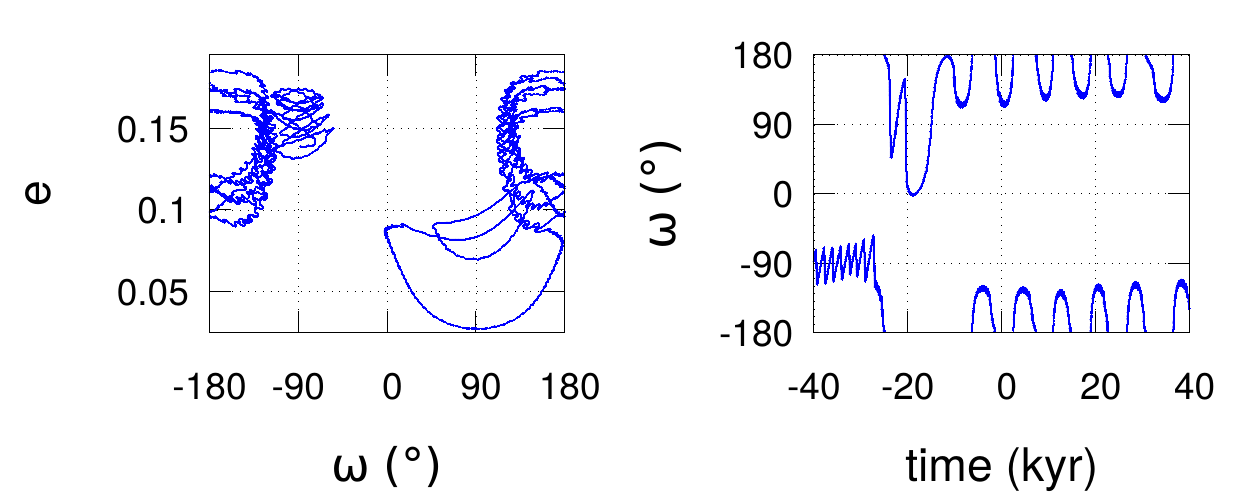}
	     \caption{Same as (a) but when all planets and asteroids, except
	     Earth, the Moon, and 2016 CO$_{246}$, are removed from the
	     calculations.}
	     \label{fig:2016CO246_onlyE}
     \end{subfigure}
	    \caption{ $e$-$\omega$ and $t$-$\omega$ graphics corresponding to
	    the $\mp$40000 yr dynamical evolution of 2016 CO$_{246}$.}
      \label{fig:2016CO246_ew}
\end{figure}

We performed additional tests for $\mp$40000 yr to see if Lidov-Kozai resonance
is the cause of the effect here, and to clarify which is the main perturber. The
results are shown in Fig.~\ref{fig:2016CO246_ew}. In Fig.~\ref{fig:2016CO246_wAll}, 
the integration of the orbit of the asteroid has been carried out using a model
Solar system that includes the eight known planets, the dwarf planet (1) Ceres,
the main-belt asteroids (2) Pallas, (4) Vesta, (10) Hygiea, (31) Euphrosyne and
the Moon, as in previous integrations.  Oscillations of $\omega $ around
90${\degr}$ in Fig.~\ref{fig:2016CO246_wAll} correspond to transitions between
the horseshoe and quasi-satellite orbit types due to the gravitational influence
of other planets (probably Jupiter or Venus) as shown in Fig.~15 of
\citet{namouni1999secular}. However, oscillations at 0${\degr}$, 180${\degr}$
and -180${\degr} $ fit the Lidov-Kozai resonance state due to the perturbation
of Earth.

In Figure \ref{fig:2016CO246_noJ}, which corresponds to the case where Jupiter
is excluded from the calculations, the number of oscillations corresponding to
orbit-type transitions decreases, although they do not disappear completely. However,
the number of oscillations around $\mp$180${\degr}$ increases. This shows that
Jupiter's perturbation cancels out the impact of the Lidov-Kozai resonance caused
by Earth.

In Figure~\ref{fig:2016CO246_noV}, when Venus is removed from the calculations,
orbit transitions appear around both $\mp$90${\degr}$. Here the main perturber
is Jupiter again. In Figure~\ref{fig:2016CO246_onlyE}, where all objects except
Earth, the Moon, 2016 CO$_{246}$ and the Sun have been removed from the
calculations, Lidov-Kozai resonance is observed between -10000 yr and 40000
yr. The oscillations are around $\mp$180${\degr}$. 

In summary, the picture emerging from the study of the dynamical behaviour of
the asteroid 2016 CO$_{246}$ indicates that this object is moving along a stable
chaotic orbit that shows brief episodes of Lidov-Kozai resonant behaviour.

\subsection{Asteroid 2017 SL$_{16}$}

\begin{figure*}
	\includegraphics[width=\textwidth]{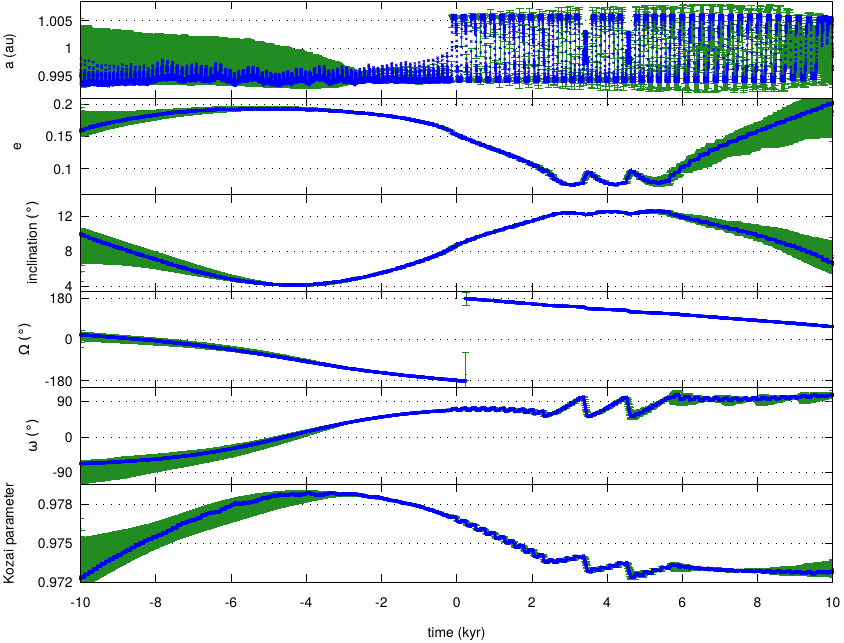}
	\caption{Average of the short-term dynamical
	evolution of 1000 clone orbits of 2017 SL$_{16}$ 
	and the nominal orbit of 2017 SL$_{16}$.
	The green curves show the average of the parameters for clone orbits, and
	the blue curves show the parameters for the nominal orbit.}
    \label{fig:2017SL16aeil_mean}
\end{figure*}

Asteroid 2017 SL$_{16}$ was discovered on 2017 September 24 by the Mt. Lemmon
survey. The orbit determination of this minor body was computed on 2019 December
12 and it is based on 58 observations for a data-arc span of 739 d.
Based on these observations, the condition code is currently $0$.  
Asteroid 2017 SL$_{16}$ is as large as 2016 CO$_{246}$ with an absolute
magnitude of 25.8 mag, which suggests a diameter in the range 12-92 m for an
assumed albedo in the range 0.60-0.01. 
The MOID for Earth is $0.0176507$ au. However, the same as 2016 CO$_{246}$, it
is not classified as PHA due to its absolute magnitude (25.8). It is currently
in the Apollo class. 

The object is moving along an asymmetrical HS-type orbit with an HS libration
period of approximately 255 yr according to the $a$, and $\lambda_r$ graphs (in
Fig.~\ref{fig2:2017SL16} panel a and c). As the horns of
the HS in the relative orbit are approximately $20{\degr}$ away from the
Earth-Sun direction, close approaches to Earth are more frequent and deeper
(see Fig.~\ref{fig2:2017SL16}, panel b) than in the case of 2016 CO$_{246}$

In Fig.~\ref{fig2:2017SL16}, $e$ decreases with small fluctuations, while $i$
increases. At the same time, the Kozai parameter and $\omega$ show almost periodic
oscillations compatible with these fluctuations. The oscillations of $\omega$
become even more significant when the orbit turns into the HS-type;
thereafter, it oscillates around $70{\degr}$.

The dynamical evolution of the semimajor axis $a$ in
Fig.~\ref{fig:2017SL16aeil_mean} shows that the nominal orbit becomes compatible
with the average of its clones after it turns into HS-type. This dynamical
behaviour lasts for about $2500$ yr. After the first transition from HS to QS
co-orbital state (about 3370 yr into the future), the average value $a$ of the
clones starts to diverge from that of the nominal orbit of 2017 SL$_{16}$.
With the second HS-QS transition (about 4517 yr into the future), the separation
between the average values of the orbital elements of the clones and the values
of the nominal orbit become significant for the values of $e$, $i$ and
$\omega$.

The dynamical behaviour of 2017 SL$_{16}$ in the interval -1000 yr -- 2300 yr
shows that $e$ and $i$ have consistent decreases/increases that keep the value
of the Kozai parameter almost constant.  However, since the moment the orbit
turns into HS-type, short-period small oscillations are observed both in
${\omega}$ and in the value of the Kozai parameter.  While $i$ and $e$ are
decreasing/increasing in the $0$ yr - $2300$ yr interval of time, they oscillate
around $70{\degr}$.
Approximately at $2300$ yr into the future, the $i$ value reaches its maximum
while the $e$ value reaches its minimum; $\omega$ begins to oscillate around
$75{\degr}$. Meanwhile, the Kozai parameter remains nearly constant. The peaks
observed at the time corresponding to the QS transitions and the long periods of
change are quite similar to those that correspond to the orbit of the minor body
2016 CO$_{246}$.

\begin{figure}
    \begin{subfigure}[b]{1.0\linewidth}
	\includegraphics[width=\linewidth,height=0.45\linewidth]{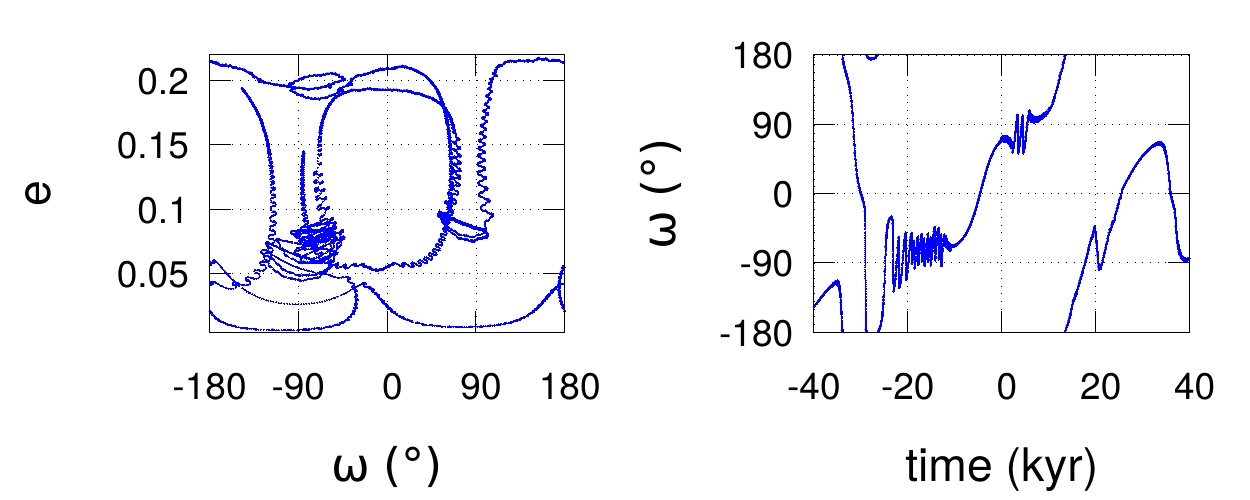}
	    \caption{Full integration for 2017 SL$_{1}$ with all planets and asteroids.}
	     \label{fig:2017SL16_wAll}
     \end{subfigure}
     \begin{subfigure}[b]{1.0\linewidth}
	\includegraphics[width=\linewidth,height=0.45\linewidth]{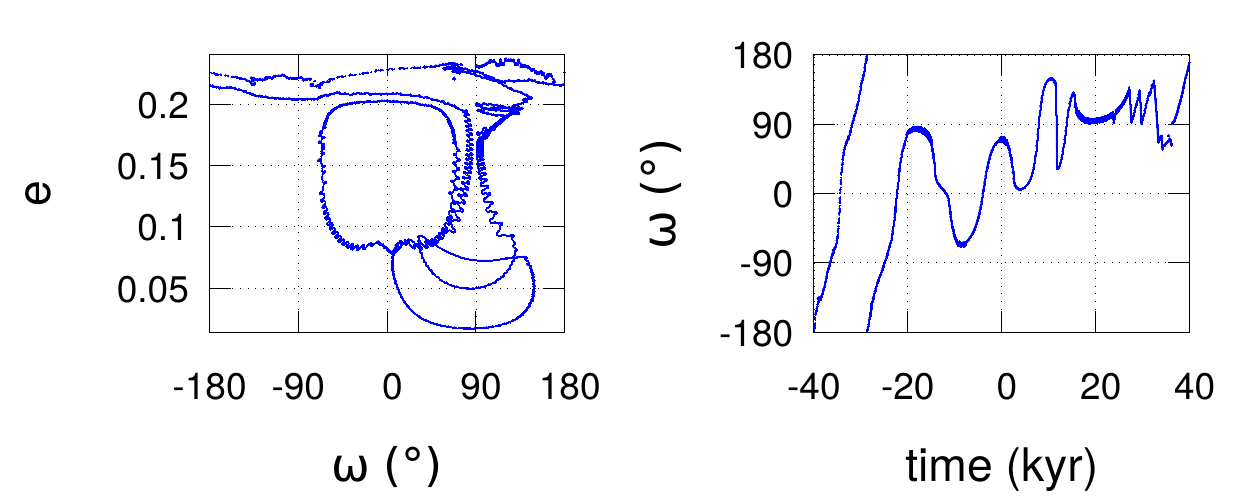}
		\caption{Same as (a) but when Jupiter is removed from the calculations.}
		 \label{fig:2017SL16_noJ}
     \end{subfigure}
     \begin{subfigure}[b]{1.0\linewidth}
	\includegraphics[width=\linewidth,height=0.45\linewidth]{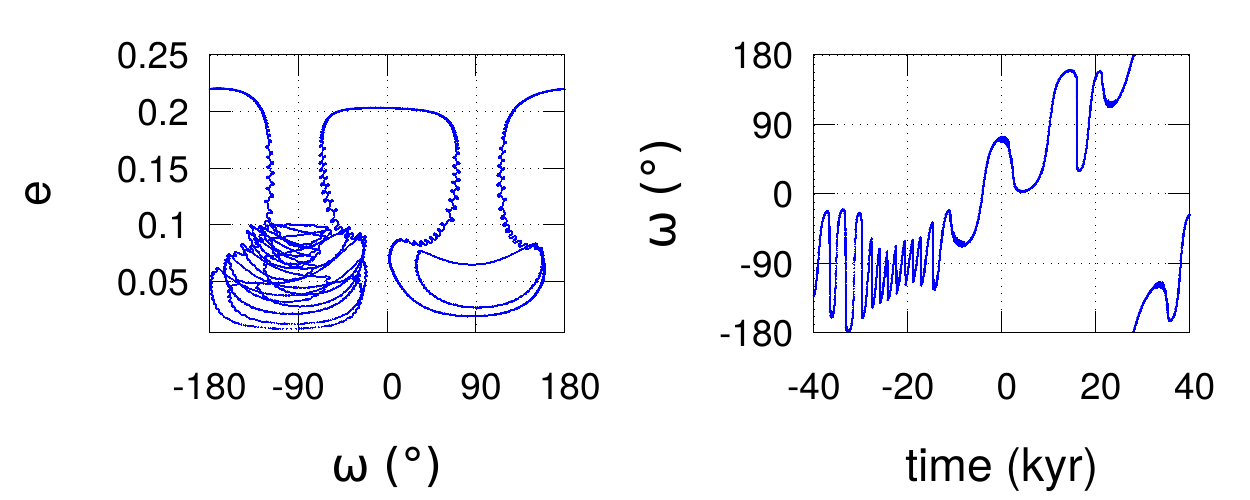}
	     \caption{Same as (a) but when Venus is removed from the
	     calculations.}
	     \label{fig:2017SL16_noV}
     \end{subfigure}
     \begin{subfigure}[b]{1.0\linewidth}
	\includegraphics[width=\linewidth,height=0.45\linewidth]{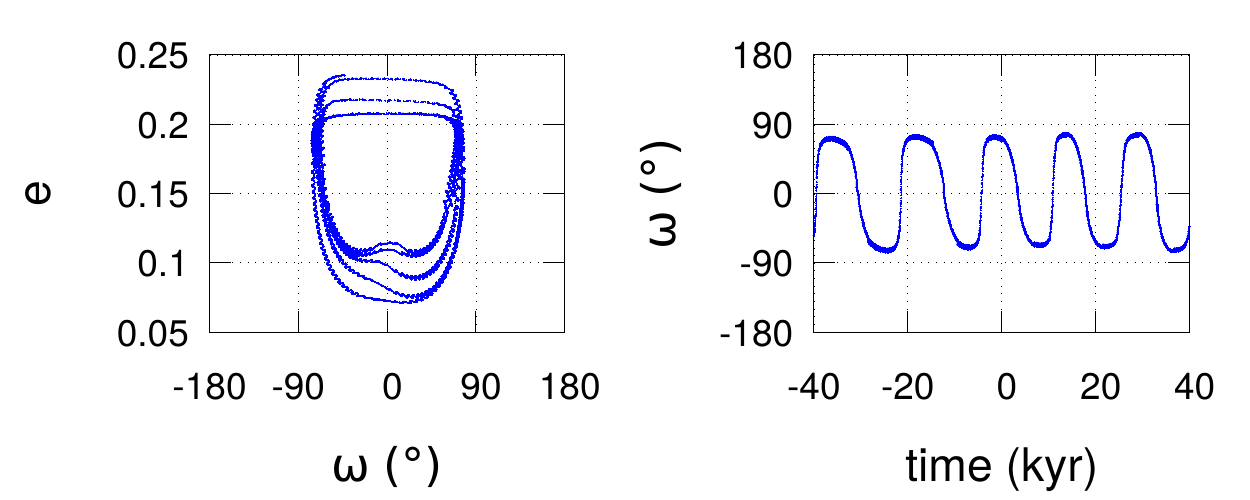}
	     \caption{Same as (a) but when all planets and asteroids, except
	     Earth, the Moon, and 2017 SL$_{16}$, are removed from the calculations.}
	     \label{fig:2017SL16_onlyE}
     \end{subfigure}
	    \caption{$e$-$\omega$ and $t$-$\omega$ graphics corresponding to
	    the $\mp$40000 yr dynamical evolution of 2017 SL$_{16}$.}
      \label{fig:2017SL16ew}
\end{figure}

Similar to the asteroid 2016 CO$_{246}$, we performed additional tests for the
asteroid 2017 SL$_{16}$ for $\mp$40000 yr range to see if the Lidov-Kozai
resonance is at work. The results can be seen in Fig.~\ref{fig:2017SL16ew}.
In Fig.~\ref{fig:2017SL16_wAll}, the integration of the orbit of the asteroid
has been carried out using a model Solar system that includes the eight known
planets, the dwarf planet (1) Ceres, the three most massive main-belt asteroids
and the Moon as in previous integrations, oscillations around -90${\degr}$ correspond
to orbit transitions between HS and QS type orbits resulting from perturbation
of other planets, just like in the case of asteroid 2016 CO$_{246}$.

It is seen in Fig.~\ref{fig:2017SL16_noJ} that in the absence of Jupiter,
oscillations at -90${\degr}$ in the past evolution disappeared completely. The
oscillations, which are indicative of the orbital transitions caused by
Jupiter's perturbation, have been replaced by oscillations around 0${\degr}$. And
these oscillations correspond to Lidov-Kozai resonance where the asteroid and
Earth are at a 1:1 mean motion resonance. In other words, the gravitational
perturbation of Jupiter disrupts the Lidov-Kozai resonance caused by Earth.

On the other hand, in the absence of Venus in the calculations
(Fig.~\ref{fig:2017SL16_noV}), the orbit transitions seen in the past evolution
of the orbit appear again for a longer period of time.
Fig.~\ref{fig:2017SL16_noJ} and Fig.~\ref{fig:2017SL16_noV} show that both
Jupiter and Venus have a similar effect on asteroid 2017 SL $_{16}$ for its
dynamical evolution in the future. In Figure~\ref{fig:2017SL16_onlyE}, when all
other planetary influences except those of Earth and the Moon are disabled, only
oscillations remain around 0${\degr}$ suggesting Lidov-Kozai resonance.

\subsection{Asteroid 2017 XQ$_{60}$}

\begin{figure*}
	\includegraphics[width=\textwidth]{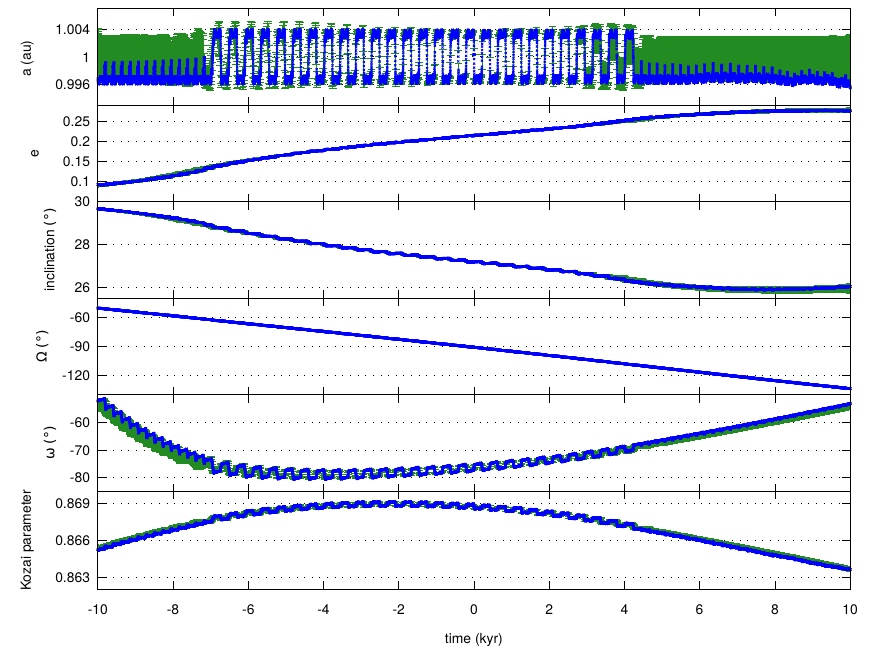}
	\caption{Average of the short-term dynamical
	evolution of 1000 clone orbits of 2017 XQ$_{60}$ and the nominal orbit
	of 2017 XQ$_{60}$. The green curves show the average of the parameters
	for clone orbits, and the blue curves show the parameters for the
	nominal orbit.}
    \label{fig:2017XQ60aeil_mean}
\end{figure*}

Asteroid 2017 XQ$_{60}$ was discovered by the Pan-STARRS survey
\citep{kaiser2004pana,kaiser2004panb} on 2017 December 13.  The orbit
determination of the body was computed on 2019 December 12 and it is based on 80
observations for a data-arc span of 738 d. The condition code, which was 8
before December 2019, was upgraded to 0 after including the observations made on
19-20-21 December 2019 in the new orbit determination.

Asteroid 2017 XQ$_{60}$'s $a$, and $\lambda_r$ graphs (Fig.~\ref{fig2:2017XQ60})
indicate an asymmetrical HS-type orbit with a libration period of approximately
410 yr.
The object probably is the largest amongst our HS co-orbital
candidates (Table~\ref{table:testasteroids}) with an absolute magnitude of
24.4 mag, which suggests a diameter in the range 23-180 m for an assumed albedo
in the range 0.60-0.01.
The inclination of the asteroid (27$\fdg$19962) has the highest value both
among candidates and among existing HS co-orbitals. The eccentricity of the
asteroid ($0.214131$) is almost as high as the eccentricities of unusual HSs
(see Tables~\ref{tab:hscoorbitals} and \ref{table:testasteroids}).
The MOID value for Earth is $0.0193109$ au. It is also in the Apollo class like
2017 SL$_{16}$, and 2016 CO$_{246}$. However, our numerical simulations point
out that it's current value of the semimajor axis is decreasing. In 2020, the semimajor
axis value is going to drop below 1 au, and the dynamical class of the asteroid
will change to Aten. It will return to the Apollo class after 211 yr of
dynamical evolution.

The nominal orbit shows asymmetrical HS characteristics between -7000 yr and about
4250 yr, and the dynamical evolution of all clones is consistent with this
behaviour  with slight differences
(Fig.~\ref{fig:2017XQ60aeil_mean}). Unlike  2016 CO$_{246}$ and 2017 SL$_{16}$,
no HS-QS transition is observed in the evolution of the orbit at any time
between -10000 yr and 10000 yr. The object remains in the 1:1 mean-motion
resonance region within this time interval for all clones. In addition, there
are no significant differences between the nominal trajectory and the averages
of the $e$, $i$, $\omega$ and $\Omega$ values for clone orbits in this time
interval.  Although $e$ and $i$ show a proportional increase/decrease between
-10000 yr and 10000 yr, no significant oscillation of $\omega$ suggesting
Lidov-Kozai resonance is observed in the same time interval.


\subsection{Asteroid 2018 PN$_{22}$}

Asteroid 2018 PN$_{22}$ was discovered on 2018 August 13 by the Pan-STARRS
survey \citep{kaiser2004pana,kaiser2004panb}. The orbit determination of this
minor body was computed on 2019 August 3 and it is based on 19 observations
for a data-arc span of 29 d. However, based on these observations, the
uncertainty of the calculated orbit is not so bad.  The condition code is
currently 3. Asteroid 2018 PN$_{22}$ is smaller than 2017 SL$_{16}$ and 2016
CO$_{246}$ with an absolute magnitude of 27.5 mag, which suggests a diameter in
the range 5-42 m for an assumed albedo in the range 0.60-0.01. The Earth MOID of
asteroid 2018 PN$_{22}$ is 0.0115656 au. The asteroid is in an Earth-like orbit with
small $e$ and $i$ values and it fits the orbital parameter space of the
dynamically cold resonant family where 0.985 au $< a <$ 1.013 au, 0 $< e <$ 0.1 
and 0${\degr}$ $< i < $ 8.56${\degr}$ defined by \citet{de2013resonant}.

According to the calculations made with the nominal orbit parameters
(Fig.~\ref{fig2:2018PN22}), the object has recently reached the nearly symmetric
HS-type orbit; it will continue moving along this type of orbit for around
175 yr. Its HS libration period is almost $125$ yr. The object may
remain longer trapped within the 1:1 mean-motion resonance with Earth or in its
immediate neighbourhood, but it does not show a stable co-orbital behaviour,
transitions out and in of the co-orbital state are occasionally observed. 

\begin{figure*}
	\includegraphics[width=\textwidth]{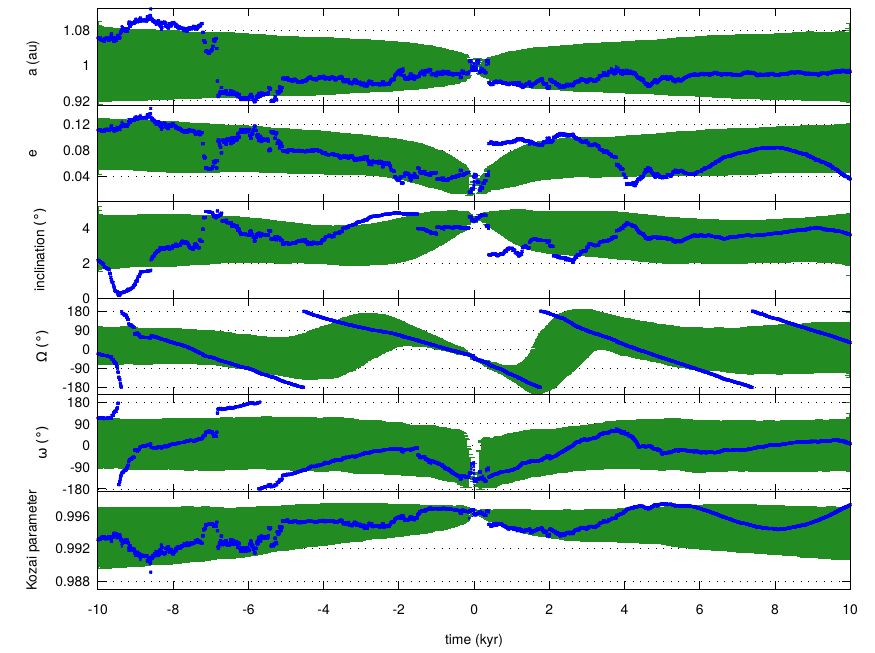}
	\caption{Average of the short-term dynamical
	evolution of 1000 clone orbits of 2018 PN$_{22}$  and the nominal orbit
	of 2018 PN$_{22}$. The green curves show the average of the parameters
	for clone orbits; the blue curves show the parameters for the nominal
	orbit.}
    \label{fig:2018PN22aeil_mean}
\end{figure*}

In this case, Fig.~\ref{fig:2018PN22aeil_mean} indicates that the averages of the
orbital parameters of the clones differ considerably from those corresponding to
the nominal orbit. During the first 60 yr of the simulations, both forward and
backward in time, all clones follow fairly similar orbits as the nominal one.
But from this point on, all orbital elements of the clones are beginning to
follow chaotic paths that diverge. Despite its small $e$ and $i$ values, 2018
PN$_{22}$ is in an unstable region of the orbital parameter space leading to a
chaotic dynamical behaviour. As stated in \citet{hollabaugh1973earth}, the
smaller the libration period, the higher the chaotic nature of the orbit.
Considering that 2018 PN$_{22}$ has the smallest libration period among our HS
candidates,  the fact that its dynamical evolution is more chaotic
than those of our other HS candidates is not surprising. The probability distribution 
that reflects this diverging behaviour is given by: 27.5 per cent of them corresponding 
to a passing object (not trapped in a 1:1 mean-motion resonance) and 72.5 per cent to
an HS co-orbital state.

In different time intervals in future and past evolutions, $e$ and $i$
oscillates, alternating high $e$ and $i$. However, it is difficult to say that
${\omega}$ also oscillates at the same intervals. This means that Lidov-Kozai
resonance is not at work, at least for the dynamical behaviour that corresponds
to the nominal orbit.

\section{Conclusions}

In this work, near-Earth asteroids (NEAs) whose semimajor axes are between 0.983
and 1.017 au have been studied; it was determined that asteroids 2016
CO$_{246}$, 2017 SL$_{16}$, 2018 PN$_{22}$, 2017 XQ$_{60}$, 2018 XW$_{2}$, 2018
AN$_{2}$ are currently asymmetrical HS co-orbitals of Earth regarding their
nominal orbital parameters. Numerical simulations were carried out for four
of these objects, namely 2016 CO$_{246}$, 2017 SL$_{16}$, 2017 XQ$_{60}$, and
2018 PN$_{22}$, selected because of their low orbit uncertainties to ensure the
reliability of the results obtained for their dynamical behaviour.  For 2016
CO$_{246}$, 2017 SL$_{16}$, and 2017 XQ$_{60}$, the dynamical evolution of all
clones is consistent with a dynamical scenario in which these bodies move along
{\it asymmetrical HS}-type orbits. The orbit of 2018
PN$_{22}$ seems to be highly more chaotic than those of 2016 CO$_{246}$, 2017
SL$_{16}$, and 2017 SL$_{16}$; asteroid 2018 PN$_{22}$ will stay less than 200
yr in its current dynamical state. However, 2018 PN$_{22}$ has a relatively high
probability (72.5\%) of moving along an HS-type orbit.

In addition, the known HS Earth co-orbitals have been briefly reviewed using
their current nominal orbital parameters. Asteroids 2015 YA and 2015 YQ$_{1}$  have
been documented as {\it asymmetrical HS} librators \citep{de2016trio}. However,
the current nominal orbit of 2015 YA will not encompass Lagrangian points L3,
L4, and L5 for about $225$ yr. After $225$ yr of dynamical evolution,
2015 YA turns into an irregular HS librator for a brief period of time. 
The current dynamical behaviour of 2015 YA seems to indicate that this object
rather tends to move along an unusual QS-type orbit than an HS-type one.
Nevertheless, according to the dynamical behaviour observed for some clones,
2015 YA may still be able to move along an HS-type orbit but this possibility
has a low probability regarding what the analysis of its current nominal orbit
indicates. A similar situation is observed for 2015 YQ$_{1}$ that is changing
its dynamical state in a relatively short time-scale. Its current nominal orbit
indicates that it is in an unusual HS co-orbital state with a high value of $e$
and an irregular dynamical behaviour of its semimajor axis $a$.

Currently, the majority of known Earth co-orbitals are moving along HS-type orbits.
HS orbits can be classified in three subgroups: unusual, nearly symmetric, and asymmetrical.
The known HS co-orbitals including the new objects studied in this work divided into the
three different categories are listed below.
{\it Unusual HSs}: (3753) Cruithne, (54509) YORP (2000 PH$_{5}$), 2015 YQ$_{1}$;
{\it nearly symmetric HSs}: 2002 AA$_{29}$, 2003 YN$_{107}$, (419624) 2010 SO$_{16}$, (454094) 2013 BS$_{45}$, 2015 SO$_{2}$, 2018 PN$_{22}$; and 
{\it asymmetrical HSs}: 2001 GO$_{2}$, 2006 JY$_{26}$, 2015
XX$_{169}$, 2016 CO$_{246}$, 2017 SL$_{16}$, 2017 XQ$_{60}$.

As in similar studies
\citep{connors2005survey,de2012dynamical,de20142013ND15,de2014asteroid,de2016trio},
non-gravitational effects as well as the gravitational influence of dwarf
planets and main-belt asteroids --except dwarf planet (1) Ceres and the main-belt asteroids (2)
Pallas, (4) Vesta, (10) Hygiea, and (31) Euphrosyne- have been neglected. The
non-inclusion of these effects has no significant influence on determining the
short-term dynamical evolution of the asteroids studied here. However, they
should be taken into account for long-term trajectory analyses and detailed
orbital stability calculations.

\section*{Acknowledgements}
The authors would like to thank the referee for his/her constructive and helpful report.
This research has been supported by the Scientific and Technological Research
Council of Turkey (TUBITAK) grand number: 118F025. The numerical calculations reported in
this paper were partly performed at TUBITAK ULAKBIM, High Performance and Grid
Computing Center (TRUBA resources). This research has made use of NASA JPL
Horizons On-Line Ephemeris. Simulations in this paper made use of the REBOUND
code which can be downloaded freely at http://github.com/hannorein/rebound.

\section*{DATA AVAILABILITY}
The data underlying this article will be shared on reasonable request to the
corresponding author.


\bibliographystyle{mnras}
\bibliography{references} 

\bsp	
\label{lastpage}
\end{document}